\documentclass[%
 reprint,
 amsmath,amssymb,
 aps, prx
]{revtex4-2}

\usepackage{graphicx}
\usepackage{dcolumn}
\usepackage{bm}
\usepackage{hyperref}
\usepackage[mathlines]{lineno}
\usepackage{bbold}
\usepackage{siunitx}
\usepackage{braket}
\usepackage{lipsum}
\usepackage{subcaption}
\usepackage[justification=raggedright]{caption}
\usepackage{multirow} 

\begin{document}
\title{Impact of static disorder on quasiparticle spectra: Debye–Waller, mean free path and potential fluctuation effects}
\author{Enrico Della Valle$^{1,2}$}
\email{enrico.della@psi.ch}
\author{Procopios Constantinou$^{1}$}
\author{Thorsten Schmitt$^{1}$}
\author{Matthias Muntwiler$^{1}$}
\author{Gabriel Aeppli$^{1,2,3}$}
\author{Vladimir N. Strocov$^{1}$}
\affiliation{$^{1}$Center for Photon Science, Paul Scherrer Institut, 5232 Villigen, Switzerland}
\affiliation{$^{2}$Department of Physics and Quantum Center, Eidgenossische Technische Hochschule Zürich, 8093 Zurich, Switzerland}
\affiliation{$^{3}$Institut de Physique, École Polytechnique Fédérale de Lausanne, 1015 Lausanne, Switzerland}

\date{\today}

\begin{abstract}
Angle-resolved photoemission spectroscopy (ARPES) is a widely used characterization technique in condensed matter physics, providing direct access to the single-electron spectral function of crystals, including their electronic band structure and Fermi surface. Measuring the band structure of novel quantum materials has been fundamentally important for determining, for example, non-trivial band topology, as in topological insulators, Weyl, and Dirac semimetals, or for identifying new classes of materials, such as altermagnets. A key challenge with these emerging quantum materials is that their initial crystalline quality is rarely optimized, which directly affects the spectra measured by ARPES. Here, we present a theoretical framework and experimental evidence addressing two common consequences of static disorder in photoemission experiments: the loss of coherent spectral weight and the broadening of spectral features. ARPES spectra can be understood as a sum of coherent and incoherent intensities, with their relative contributions controlled by atomic disorder. Under thermal disorder, the coherent intensity is exponentially suppressed as temperature increases, a phenomenon analogous to the Debye–Waller factor in diffraction, where Bragg peaks diminish in favor of diffuse scattering as disorder increases.
In this work, we report a soft X-ray study of the deliberately disordered (via Ar ion sputtering) InAs(110) surface, characterized both by scanning tunneling microscopy (STM) and low energy electron diffraction (LEED). We introduce a new framework that enables quantification of coherent photoemission intensity loss with increasing disorder, allowing both thermal and static disorder to be treated within a unified approach. Additionally, we identify a second major effect of disorder beyond lifetime broadening: inhomogeneous spectral broadening arising from local potential fluctuations. We show that such fluctuations increase the linewidths of the spectra of localized and delocalized states, and contribute to the suppression of ARPES intensity from states near the Fermi level. The concepts and analysis methods presented should make ARPES useful for direct diagnosis of disorder effects on electronic states, for science as well as engineering.  
\end{abstract}

\maketitle


\section{Introduction}
Angle-resolved photoemission spectroscopy (ARPES) accesses a plethora of information about the occupied electronic states of crystalline solids, including single-electron spectral function, band structure and Fermi surface \cite{Strocov2019,Damascelli2003}. 
The surface/bulk sensitivity can be actively tuned by the photon energy, $h\nu$. In particular, the use of soft X-rays, where $h\nu$ spans from 300 eV up to 2000 eV, rather than extreme ultraviolet (EUV) radiation, enhances the bulk sensitivity. This generally increases the inelastic mean free path  (IMFP), with typical values of 1-2 nm restricted primarily by electron-electron interactions of the photoexcited electrons, which are in the final states of the photoemission process \cite{Powell2020}. The consequent increase of bulk sensitivity allows soft X-ray ARPES (SX-ARPES) to optimally probe the electronic structure of three-dimensional (3D) materials and buried interfaces \cite{Strocov2019,Kobayashi2012,Strocov2014,Woerle2017}.

Of special interest for both conventional and ``quantum'' technologies are heterostructures, which incorporate quantum wells \cite{Constantinou2023}, as well as metal-semiconductor \cite{Strocov2019} and superconductor-semiconductor interfaces \cite{Schuwalow2021,Yu2021,Yang2022} for a variety of applications.
For example, the narrow-gap semiconductor InAs coated with superconducting aluminium has attracted attention as a host for Majorana zero modes, due to appreciable spin-orbit coupling and a large Landé g factor \cite{Msft2025,Lutchyn2018,Sarma2015}. 
Disorder has detrimental effects on the robustness of these states \cite{Anderson1959}, and here we use ARPES to directly measure its impact.

Generally, ARPES spectra can be decomposed into two components: (i) a sharp, momentum-dependent coherent component ($I_{coh}$) that originates from Bloch electrons in the unperturbed crystal structure; (ii) a (usually) broad incoherent component ($I_{inc}$), which originates from thermal vibrations, elastic scattering from defects, and many-body physics. 
A theory that is well established in reproducing ARPES spectral intensity is the one-step model introduced by Pendry \cite{Pendry1976}, Feibelman \cite{Feibelman1974}, Mahan \cite{Mahan1970}, Schaich and Ashcroft \cite{Schaich1970}, and used to successfully model many condensed matter systems, with $h\nu$ ranging between a few eV and $10 \text{ keV}$ \cite{Braun2018}. 
In this framework, all electrons in the unperturbed crystal are described by Bloch waves and the states of the photoexcited electrons are identical to time-reversed low-energy electron diffraction (LEED) states \cite{Inglesfield1992}. Consequently, it is not surprising that quasiparticle peaks in ARPES spectra behave similarly  to Bragg peaks in LEED. For example, thermal vibrations and uncorrelated defects in LEED transfer intensity from sharp diffraction spots to a background homogeneous in momentum space ($\textbf{k}$-space) \cite{Henzler1984}. In ARPES, we expect a similar transfer from quasiparticle peaks in $I_{coh}$ to the $I_{inc}$ component. This has already been established for thermal disorder in the seminal work of Schevchik \cite{Shevchik1977,Shevchik1978,Shevchik1979}, who showed that the relation between $I_{coh}$ and $I_{inc}$ can be described in terms of the Debye–Waller factor (DWF), as discussed in Sect.~\ref{Models: The coherence residue}. However, no progress has been made thus far in developing a similar intuitive description for static disorder. We show here that an extended version of the approach of Schvchik \cite{Shevchik1977,Shevchik1978,Shevchik1979} to the DWF explains the impact of deliberately introduced static disorder in a model system, namely InAs(110) damaged by sputtering.

Additionally, static disorder in the form of charged defects in insulators and weakly doped semiconductors induces spectral broadening through two distinct mechanisms: (i) elastic scattering of initial states by the disorder potential and (ii) broadening caused by spatial variations in band bending \cite{Schofield2013}, referred to as inhomogeneous broadening in this manuscript. The first mechanism, discussed in Sect.~\ref{Results: Momentum broadening}, is characterized by a Lorentzian lineshape of the quasiparticle peaks and depends on the spacing between defects, scaling as $c_d^{1/3}$ in 3D, where $c_d$ is the defect concentration. The second mechanism, discussed in Sect.~\ref{Results: Inhomogeneous broadening}, becomes relevant in insulators and semiconductors when the density of compensated charged defects in the crystal is higher than the electron (or equivalently, hole) density. In this case, Coulomb defects are only partially screened, leading to potential fluctuations, also referred to as variable band bending. In particular, this broadening mechanism becomes significant when the screening length of the system is comparable to the inter-defect distance. In such cases, the potential fluctuations extend throughout the whole crystal volume. When spectra are measured using a beam spot size several orders of magnitude larger than the screening length, the resulting spectra represent an average over all potential configurations. This mechanism, unlike (i), is characterized by Gaussian lineshapes of the quasiparticle peaks and scales as $c_d^{1/2}$. These fluctuations lead to inhomogeneous broadening, affecting not only tightly bound core-level spectra but also the spectra of delocalized valence and conduction states along the energy coordinate. This mechanism has been suggested in previous theoretical studies~\cite{Chen2009} to explain the large spectral linewidths observed in high-$T_c$ cuprates in the insulating phase, as measured using ARPES in Refs.\cite{Wells1995,Pothuizen1997,Yoshida2003,Yoshida2007}. While different explanations have been proposed for these observations, such as in Ref.\cite{Shen2004}, we provide compelling evidence in this paper for the universal nature of the inhomogeneous broadening mechanism in insulators and weakly doped semiconductors.
In fact, our experimental data show that the linewidths of the valence bands and the In 4d core levels exhibit a consistent square-root dependence on the disorder concentration. A direct consequence of the fluctuating potential in the presence of a finite electron density is that, when the ratio between the bandwidth of the conduction electrons and the disorder strength is smaller than unity, electrons form disconnected puddles. These puddles, in the low-temperature regime, manifest in transport measurements with the typical conductivity of Efros-Shklovskii variable range hopping \cite{Shklovskii72, Shklovskii1984}. This phenomenon has no intrinsic energy scale, as it depends on the aforementioned ratio. In particular, it is also highly relevant for the quantum applications mentioned above. In our experiment, variable band bending leads to a measurable decrease in the intensity of the conduction states.

\begin{figure*}
    \centering
    \includegraphics[width=\linewidth]{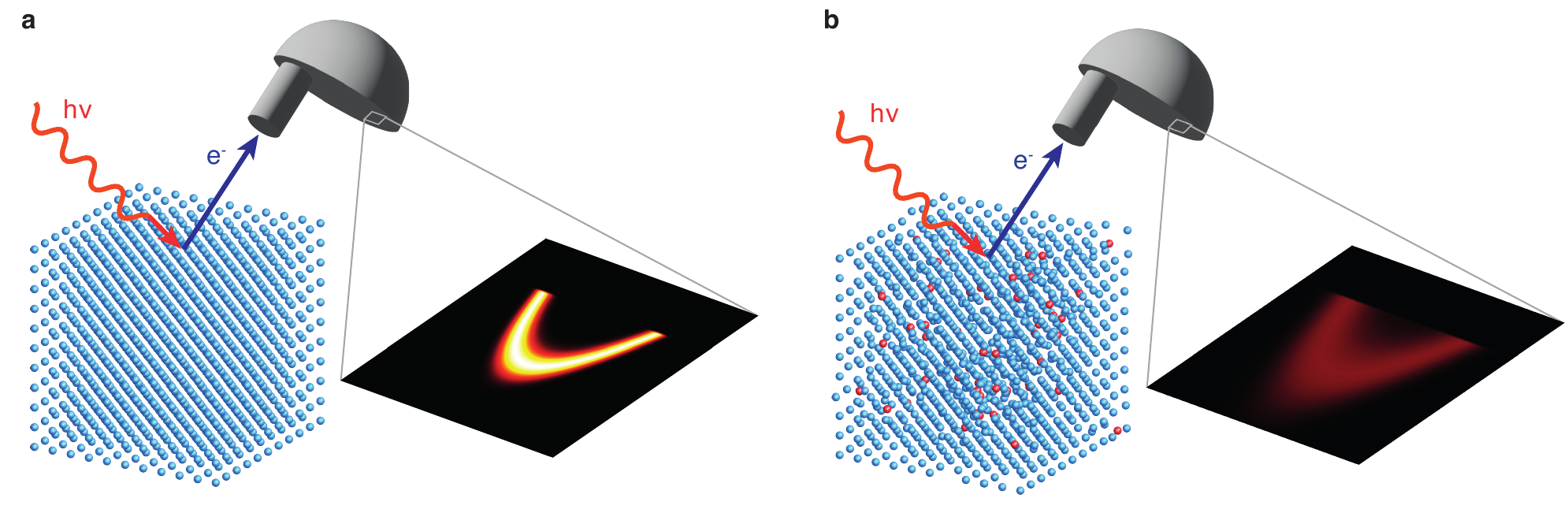}
\caption{(a) Schematic representation of an ARPES experiment performed on a well-ordered crystal (left), along with the corresponding schematic of a parabolic spectral function. (b) Schematic representation of an ARPES experiment performed on a disordered crystal, where the coherent intensity of the spectra is weakened and the lineshapes are broadened.}
\label{fig1:principle}
\end{figure*}

Therefore, as shown schematically in Fig.~\ref{fig1:principle}, the two main impacts of static disorder on ARPES spectra measured on crystals are the transfer of intensity from the coherent component, $I_{\text{coh}}$, to the incoherent component, $I_{\text{inc}}$, in exact analogy to what happens in crystallography, where Bragg peaks persist while losing intensity to diffuse scattering as disorder is increased; and spectral broadening, resulting from a reduced lifetime of initial states due to electron-defect scattering, as well as from inhomogeneous broadening caused by spatial fluctuations in the potential.

\section{Methods}
\label{Experimental procedures}
The ARPES measurements were performed at the soft-X-ray ARPES facility \cite{Strocov2014} of the ADRESS beamline \cite{Strocov_2010, Strocov2013} of the Swiss Light Source, PSI, Switzerland. 
The sample temperature was constant at  $\text{T} = 12 \text{ K}$, which
minimizes the loss of spectral weight due to thermal disorder \cite{Braun2013}, and as discussed in Sect.~\ref{Models: The coherence residue}, restricts the main visible effect to a mere prefactor weighting the coherent quasiparticle intensity.

We used vertically polarized light (p-pol in the actual experimental geometry) at three different values of $h \nu$: $517 \text{ eV}$, $815 \text{ eV}$ and $1159 \text{ eV}$. 
The $k_z$ component of the final wave vector can be calculated from the experimental geometry, the energy, and the momentum of the photons \cite{Strocov2013}. 
The above photon energies were selected so that the calculated k$_z$ would correspond to the $\Gamma$-point, at $k_x$, $k_y=0$.
Furthermore, the $h \nu$ and the (110) orientation of the InAs surface allowed us to measure the band structure along the $\Gamma$KX high-symmetry direction. 
This is the ideal direction to probe the valence bands due to the relatively large momentum-dependent separation of the heavy-hole (HH), light-hole (LH), and split-off (SO) bands, as shown in Fig.~\ref{fig2}(a).

The samples used for SX-ARPES, scanning tunneling microscopy (STM), and LEED experiments, with dimensions of $4 \times 4 \text{ mm}^2$, were cut from an undoped InAs(110) wafer grown by Wafer Technology LTD. The samples were mounted on copper Omicron plates using epoxy (EPO-TEK\textsuperscript{\textregistered} H21D) and subsequently cured at $150^\circ \mathrm{C}$ for 5 minutes. Next, a metal cleavage post was attached to the top surface of the samples using the same epoxy and curing procedure. After insertion into ultrahigh vacuum, the samples were thermalized at $10$ K for the SX-ARPES experiment and $80$ K for STM and LEED, and subsequently cleaved by removal of the cleavage post. To artificially introduce defects into the crystal structure, InAs(110) substrates were cumulatively sputtered at room temperature with Ar ions in steps of $\approx 20 \text{ s}$. 
The incident ion energy was $3 \text{ keV}$, and the partial pressure of Ar was $5\times 10^{-8} \text{ mbar}$.
Thermal annealing was not performed to ensure that the sputter-induced disorder persisted. After each sputter cycle, SX-ARPES measurements with equal integration time were carried out to record the evolution of the spectra as a function of the Ar$^+$ ion dose.
To assess the impact of sputtering-induced disorder, we analyzed the lineshapes of the quasiparticle peaks in the valence band region. Our approach involved separation, as explained in Appendix~\ref{App: background subtraction}, of the coherent signal from the incoherent background. From the processed spectrum measured at $517 \text{ eV}$ ($815 \text{ eV}$, $1159 \text{ eV}$), we extracted $20$ ($15$, $10$) MDCs within the binding energy range of $[-4 \text{ eV}, -2\text{ eV}]$ ($[-3.5 \text{ eV}, -2\text{ eV}]$, $[-3.5 \text{ eV}, -2.5\text{ eV}]$), using an integration window of $\pm 0.03 \text{ eV}$. Each coherent peak within these MDCs was fitted using the procedure described in Appendix~\ref{App: coherence residue and linewidths}. This step was repeated for data obtained after each sputtering cycle, normalizing each peak area to the initial peak area before sputtering. The coherent residue for a given Ar$^+$ dose was calculated by averaging the normalized peak areas across all MDCs. In addition, as further detailed in Appendix~\ref{App: coherence residue and linewidths}, we analyzed the FWHM of the peaks to evaluate the broadening effects. Another analysis was performed on the conduction band region measured at $517 \text{ eV}$ and highlighted by the expanded boxes in Figs.~\ref{fig2}(a)–(c). The size of these boxes in $k_x$ is $[-0.1 \text{ \AA}^{-1}, 0.1\text{ \AA}^{-1}]$, while along $E_b$ it is $[-0.2 \text{ eV}, 0.1\text{ eV}]$. To extract the total coherent intensity from these regions, we integrated over the $E_b$ coordinate, then subtracted a linear background, and finally fitted the remaining peak with a Gaussian.
To characterise directly the nature and density of the sputter-induced defects which are responsible for the changes in the ARPES data, we performed STM and LEED at the PEARL Beamline of the Swiss Light Source, Paul Scherrer Institut. The STM data were acquired using an Omicron LT STM. The tip, made of PtIr (90\% Pt, 10\% Ir), was prepared by cutting a PtIr wire, then conditioned in ultrahigh vacuum (UHV) by repeated voltage pulses and controlled tip crashes into a clean Au(111) crystal surface until stable and high-resolution images without tip artifacts were achieved. The STM images were recorded at 4.4 K.

\section{Results and discussion}
\subsection{SX-ARPES measurements}
\label{SX-ARPES measurements}
\begin{figure*}
    \centering
    \includegraphics[width=\linewidth]{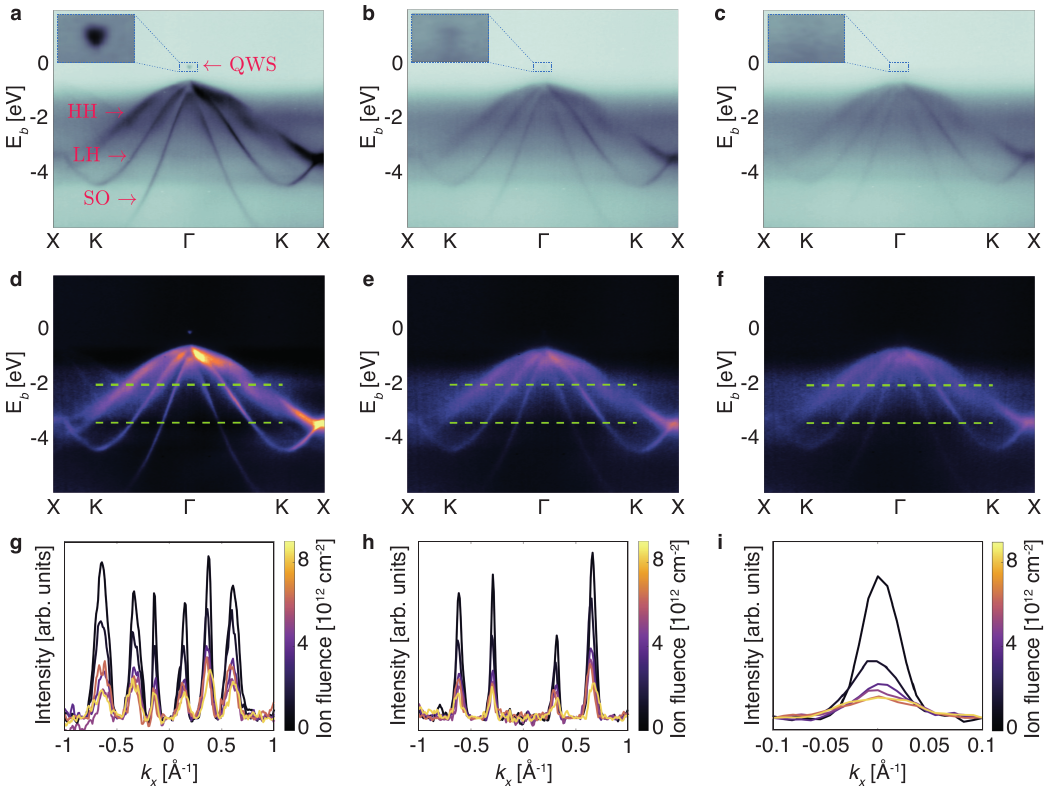}
\caption{(a)–(c) Band structure of InAs along $\Gamma$KX measured at $h\nu = 517 \text{ eV}$. (a)–(c) Total ARPES intensity shown in ascending order of Ar$^+$ fluence: $D=0$ cm$^{-2}$, $D=4.5\times 10^{12}$ cm$^{-2}$, $D=9\times 10^{12}$ cm$^{-2}$. In the upper left corner, an expanded image of the quantum well states (QWS) is shown for all three spectra. The heavy-hole (HH), light-hole (LH), split-off (SO), and QWS bands are individually identified in (a). (d)–(f) Same data as above, after subtraction of the $\textbf{k}$-integrated background. The colored lines indicate the positions where the MDCs shown in (g) and (h) have been extracted. (g)–(h) MDCs measured after increasing Ar$^+$ doses, as indicated by the colorbar on the right. (i) Intensity of the expanded boxes in (a)–(c) after integration over $E_b$ and subtraction of a linear background.
}
\label{fig2}
\end{figure*}

Fig.~\ref{fig2} presents the soft X-ray angle-resolved photoemission spectroscopy (SX-ARPES) data along the $\Gamma$KX high-symmetry line at $h\nu = 517 \text{ eV}$, with increasing Ar$^+$ fluences. In Fig.\ref{fig2}(a), we clearly identify the HH, LH, and SO valence bands. Figs.~\ref{fig2}(a-c) show the ARPES data in ascending order of the Ar$^+$ dose, demonstrating that the coherent component of the valence bands remains robust against sputtering, as HH, LH, and SO can still be resolved in Fig.~\ref{fig2}(c).

However, with increasing Ar$^+$ doses, the relative intensity of the coherent component decreases, and a momentum-independent intensity emerges across the entire Brillouin zone. The interplay between coherent and incoherent components depends critically on the degree of static disorder introduced by ion bombardment. We explain this phenomenon by introducing the coherence residue $F_{\text{C}}$, as discussed in Sect.~\ref{Models: The coherence residue}.

In Figs.~\ref{fig2}(d)-(f), we present the ARPES data after subtracting the $\textbf{k}$-integrated background. For the cleaved surface shown in Figs.\ref{fig2}(a) and (d), we observe a pocket occupied by quantum well states (QWS) originating from the conduction band at the $\Gamma$ point. These states disperse just below  $E_{\text{F}}$ and approximately $0.4 \text{ eV}$ above the valence band maximum. They are induced by defects, which act like donors, creating a charge accumulation in the vicinity of the InAs(110) surface \cite{Olsson1996}. As shown in the expanded images in the upper left corner of Figs.~\ref{fig2}(a)–(c), the QWS intensities appear to decrease at a faster rate with disorder compared to the valence band signal, which also displays a slight broadening with increased sputtering. Sects.~\ref{Results: Depletion of coherent VB intensity} and \ref{Results: VB spectral broadening} present a quantitative analysis.
\subsection{Spatial distribution of defects}
\label{Sec: Depth distribution of defects}

\begin{figure*}
    \centering
    \includegraphics[width=1\linewidth]{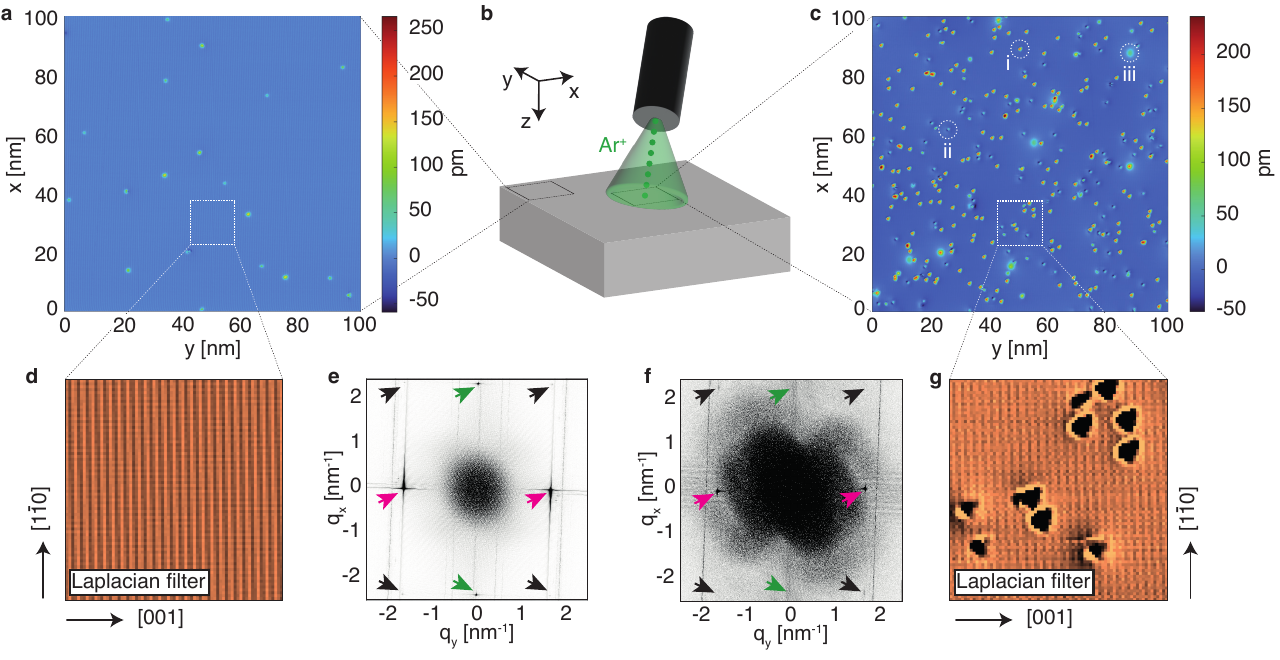}
\caption{Scanning tunneling microscope image of a $100 \times 100 \text{ nm}^2$ region of an InAs(110) surface (a) after cleavage and (c) after being subjected to a $D = 3 \times 10^{12}\text{ cm}^{-2}$ ion dose, acquired with $I = 100 \text{ pA}$ and $V = 0.4\,\text{ V}$. (b) Schematic representation of an InAs crystal being sputtered with Ar$^+$ ions. (d) and (g) expanded images of the $15\times 15\, \text{ nm}^2$ squares shown in (a) and (c) and after applying a Laplacian filter to enhance edges. (e) and (f) Absolute value of the Fourier transform of (a) and (c), respectively. The arrows indicate the Bragg peaks visible on both the pristine and disordered surfaces.} 
\label{fig3}
\end{figure*}

\begin{figure*}
\centering
\includegraphics[width=1\linewidth]{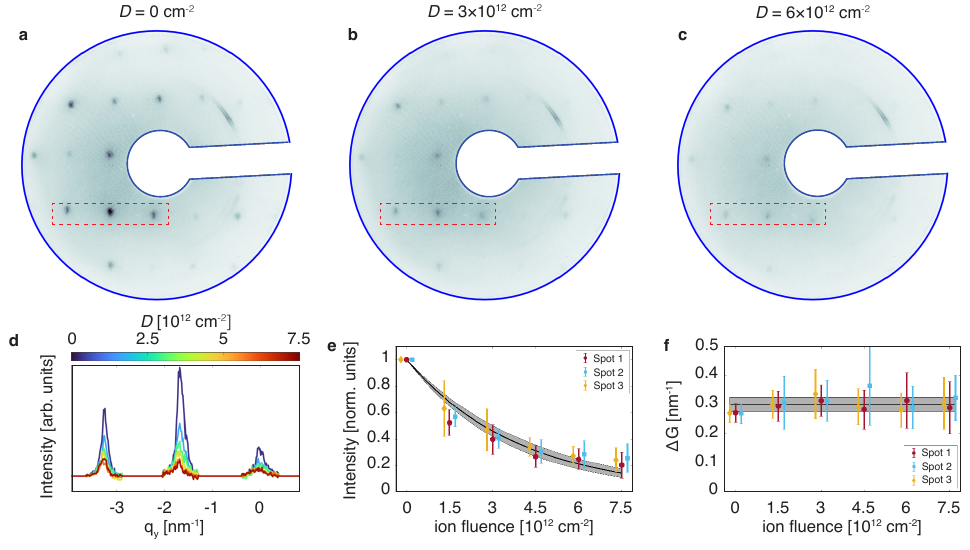}
\caption{LEED patterns measured at 100 eV kinetic energy on the same \textit{in-situ} cleaved InAs(110) surface: (a) after cleaving; (b) after Ar sputtering with a dose of $D = 3 \times 10^{12}$ cm$^{-2}$; and (c) after $D = 6 \times 10^{12}$ cm$^{-2}$. All intensities have been normalized to the maximum intensity of the freshly cleaved sample. (d) Intensity profiles extracted from the highlighted regions in (a)–(c), after linear background subtraction. (e, f) Peak intensities and FWHM values of all clearly resolved peaks at 100 eV, obtained via Gaussian fitting. The three different colors indicate data points measured at equal disorder on three distinct spots of the surface, spatially separated by 500~$\mu$m.}
\label{fig4}
\end{figure*}
Understanding the spatial distribution of sputter-induced defects is crucial for accurately interpreting the ARPES data. In this experiment, we employ a very low ion flux, several orders of magnitude lower than what is typically used for semiconductor surface preparation. This choice is motivated by the aim of preserving surface crystallinity, to investigate the effects of point defects on the momentum-resolved one-electron spectral function. Simultaneously, our method of introducing defects in a controlled manner also provides valuable insights into the ion sputtering technique itself, which is one of the most widely used surface preparation methods in semiconductor research and industry. The effects of individual ions on semiconductor surfaces have been poorly explored experimentally, resulting in only a few published papers such as Refs.~\cite{Costantini2001,Stoian2012}.

To investigate the in-plane distribution of defects, we performed both STM and LEED on \textit{in-situ} cleaved InAs(110) surfaces, before and after sputtering, to image defects in both real and momentum space. For the STM measurements, we examined a clean surface and the same surface after a total ion dose of $3 \times 10^{12}\text{ ions }\text{ cm}^{-2}$, using the conditions described in Sect.\ref{Experimental procedures}. The resultant topographic images, measured at $I=100 \text{ pA}$ and $V=0.4\text{ V}$, are presented in Figs.\ref{fig3}(a) and (c). Figure~\ref{fig3}(a) shows a $100\times100$ nm$^{2}$ region of the freshly cleaved InAs(110) surface, revealing fewer than 20 defects. The origin of these defects is attributed to the cleavage procedure. Figure~\ref{fig3}(c) illustrates the topography of the surface after sputtering. In particular, this image reveals a significantly higher concentration of defects than the non-sputtered surface. Most importantly, however, the STM image demonstrates that the underlying blue region, outlined by the $15\times15$ nm$^{2}$ dashed square and more clearly visible in the magnified view in Fig.\ref{fig3}(g), preserves long-range order. Furthermore, the absolute value of the Fourier transform of the STM image in Fig.~\ref{fig3} reveals Bragg peaks at $q_x = 2.37 \text{ nm}^{-1}$ and $q_y = 1.65 \text{ nm}^{-1}$, which correspond to the expected lattice constants $a_{(1\bar{1}0)} = a_{(001)}/\sqrt{2} \approx 0.42$ nm, and $a_{(001)} \approx 0.61$ nm, along the $x$- and $y$-directions, respectively. This confirms that the overall surface crystallinity is preserved.

Regarding the type of defects induced by Ar sputtering, the red features marked by circle i in Fig.\ref{fig3}(c), and the half-dark, half-bright blue regions marked by circle ii in Fig.\ref{fig3}(c), indicate spots where Ar ions have penetrated the crystal surface. These defects are more clearly visible in the magnified and Laplace-filtered image in Fig.\ref{fig3}(g). The difference between i and ii is attributed to the different charge states of the defects. It is also worth noting that these types of defects are absent on the pristine surface, confirming that they are related to the sputtering process. Additionally, the feature indicated by the dashed circle iii in Fig.\ref{fig3}(c) displays bright blue, blurred circular areas, which most likely correspond to sputter-induced charged defects located beneath the surface. Finally, as discussed in Sect.\ref{Models: The coherence residue}, the key to applying the Debye–Waller picture to static disorder is the notion that point defects, in this case generated by Ar$^+$ bombardment, induce distortions in the surrounding lattice. However, these displacements must occur on a length scale comparable to the wavelength of the final states in photoemission, approximately $\sim 0.1 \text{ \AA}$ in the soft X-ray regime. This scale cannot be directly resolved in real space in the data shown in Fig.\ref{fig3}, although the reduction in Bragg peak intensity in the Fourier transform shown in Fig.\ref{fig3}(f), compared to Fig.\ref{fig3}(e), already indicates the static Debye–Waller behavior discussed in Sect.\ref{Models: Effects of disorder on ARPES spectra}.

A more direct technique to examine the impact of defects on the crystal lattice at the surface in momentum space is LEED. LEED patterns were measured on freshly cleaved InAs(110) surfaces before and after Ar$^+$ sputtering, using the procedure described in Sect.\ref{Experimental procedures}. The results measured at an electron kinetic energy of 100 eV for three different defect concentrations are presented in Fig.~\ref{fig4}(a)–(c). Comparing the leftmost and rightmost patterns, we observe that the coherent diffraction peaks diminish in favor of an increasingly diffuse background. This behavior, briefly mentioned above, once again exemplifies the static Debye–Waller effect, which is a central focus of this work. From the same dataset, measured at six different defect concentrations, we extracted intensity profiles across all well-resolved peaks along the radial direction. After subtracting a linear background, we fitted Gaussian functions to extract both the peak area and the FWHM. Fig.\ref{fig4}(e) displays the peak area normalized to the area before sputtering, while Fig.\ref{fig4}(f) shows $\Delta G$, defined as the FWHM of the peaks in the radial direction. The peak intensity was fitted with an exponential function, while for $\Delta G$ we plot the mean value and the corresponding standard deviation. The mechanism responsible for the depletion of the coherent Bragg peak intensity, as previously mentioned and thoroughly discussed in Sect.~\ref{Models: The coherence residue}, also accounts for the reduction in ARPES coherent intensity. Meanwhile, the peak linewidths show no clear trend with increasing disorder, with any broadening remaining below 0.05 nm$^{-1}$. Two mechanisms could explain a possible broadening: a reduction in long-range order and microstrain induced by defects, as schematically illustrated in Fig.~\ref{figA3}. However, STM data and its Fourier transform, shown in Fig.\ref{fig3}(c) and (f), indicate that long-range coherence is preserved over at least 100 nm, implying a maximal contribution to broadening of $\sim 0.01$ nm$^{-1}$, thereby ruling out this mechanism as dominant. The results in Fig.~\ref{fig4}(f) therefore represent an upper bound for the displacement field around each point defect. The correlated atomic displacements give rise to a finite distribution of diffraction angles around the nominal Bragg condition, resulting in peak broadening. An upper bound for the magnitude of the displacement field can be estimated from the difference between $a_{(001)}(\Delta G/G)$ and the corresponding quantity calculated using $\Delta G$ plus its standard deviation, yielding a maximal correlated displacement of approximately $0.1$ \AA, consistent with the earlier discussion.

Finally, in Fig.\ref{figA2} of Appendix\ref{App: SRIM}, we present simulations of the defect depth distribution based on the Stopping and Range of Ions in Matter (SRIM) software~\cite{Ziegler2010}. These simulations are used to estimate the effective defect concentration corresponding to the three different probing depths achieved with varying photon energies. This concentration, in turn, is used to quantify the magnitude of the atomic displacement field extracted from the ARPES experimental data.

\subsection{Effects of disorder on ARPES spectra}
\label{Models: Effects of disorder on ARPES spectra}
There are three principal effects of disorder on the one-electron Green’s function as measured by ARPES which also includes final state effects: (1) a reduction in the amplitude of the quasiparticle peaks due to elastic scattering of photoelectrons by defects and the atomic displacements associated with the disorder, (2) a broadening of the peaks in momentum space due to the finite  mean free path introduced by disorder for the quasiparticles existing at equilibrium within the solid under investigation, and (3) inhomogeneous broadening in energy due to the unscreened Coulomb potential of the defects. In the remainder of this paper, we discuss (1) first, followed by (2) and (3) in turn.

\subsubsection{The coherence residue}
\label{Models: The coherence residue}
Effect (1) can be  described by an extended version of DWF.
Originally, this theory was developed to describe the effects of temperature on X-ray diffraction measurements \cite{Debye1913}, and was subsequently extended to photoemission \cite{Shevchik1977,Shevchik1978,Shevchik1979}. 
At finite temperatures, the atoms that constitute the crystal vibrate around their ideal lattice positions. 
The fact that during the photoexcitation process the atoms are not localized at their perfect lattice sites creates a random phase mismatch between the initial and final states. 
This has severe consequences for the total coherent intensity $I_{coh}$. 
In fact, depending on both the temperature and photoelectron kinetic energy, the phase mismatch between the initial and final states can gain importance and eventually deplete a large portion of the coherent intensity. 
This depletion can be quantified via the DWF, which is normally interpreted as the fraction of coherent intensity that reaches the detector \cite{Hussain1980,White1986}: 
\begin{equation}
    \text{DWF} = e^{-\Delta K^2 \langle u_T^2 \rangle}.
    \label{True DWF}
\end{equation}
In Eq.~\ref{True DWF}, $\Delta K$ is the difference between the wave vector of the initial and final states, and $\langle u_T^2 \rangle$ is the average square displacement of the atoms due to thermal vibrations. 
The former quantity, $\Delta K$, depends on the photoelectron kinetic energy and, in turn, $h\nu$. 
In particular, the higher $h\nu$, the more important the phase mismatch becomes and, consequently, the smaller the portion of the coherent intensity. 
In the usual Debye–Waller picture, $\langle u_T^2 \rangle$ is proportional to temperature.
In this paper, we extend the DWF to the case of static disorder. This step is also motivated by the successful treatment of thermal fluctuations within the one-step model \cite{Nicola__2018}, more precisely using the Coherent Potential Approximation (CPA) \cite{Soven_1967} within the multiple scattering theory or the KKR formalism. Here, the thermal fluctuations of the atoms are approximated by a finite number of static displacement vectors. Furthermore, the sum of the squares of these vectors, weighted with the probability of such a displacement, is forced to be equal to $\langle u_T^2 \rangle$ \cite{MS_2018}. An additional motivation comes from the schematics of a disordered crystal shown in Fig.~\ref{figA3} of Sect.~\ref{App: coherence residue}, corroborated by the LEED data shown in Fig.~\ref{fig4}(i). When the defect concentration is high, certain atoms find themselves displaced from their ideal lattice positions due to strain generated by nearby defects. This scenario resembles the atomic displacements associated with thermal disorder, where all atoms are displaced (quasi-)randomly from their ideal lattice sites. To understand how static disorder, which results from collision cascades of incident ions, affects the ARPES spectral intensity we introduce the concept of an averaged square displacement, $\langle u_d^2 \rangle$. This measure is influenced by the defect concentration $c_d$, and is incorporated into the Debye–Waller framework.

We begin by demonstrating that the mere creation of vacancies or interstitials in absence of a displacement field has minimal effects on the coherent photoemission intensity. In Appendix~\ref{App: coherence residue}, we consider a hypothetical scenario where $M$ vacancies and $M$ interstitials are introduced into an otherwise perfect crystal composed of $N$ atoms. In particular, this means that the additional displacement of surrounding atoms is totally neglected. Our results reveal that in this idealized case, the coherent intensity is reduced by approximately $\left| 1-M/N\right|^2$. This implies that even with a high vacancy concentration of $1\%$, the coherent intensity only decreases by about $2\%$. This observation underscores the importance of considering the secondary static atomic displacements caused by vacancies and interstitials, as depicted in Fig.~\ref{figA3}, to account for the significant reduction in coherent intensity observed at defect concentrations on the order of one percent or less. To quantify these secondary displacements, we calculate the average squared displacement as $\langle u_d^2 \rangle = \frac{1}{N} \sum_{i=1}^{N} u_i^2 = \frac{1}{N} \sum_{i=1}^{M} u_{sd}^2 = c^0_d u_{sd}^2$. Here, $c^0_d$ denotes the relative defect concentration, and $u_{sd}^2 = Z_{nn} u_{nn}^2 + Z_{nnn} u_{nnn}^2 + \ldots$ represents the total squared displacement caused by a single defect (sd), equal to the sum of the squared displacements affecting nearest neighbors ($Z_{nn}$), next-nearest neighbors ($Z_{nnn}$), and so on.

\begin{equation}
F_{\text{C}} \equiv e^{-\Delta K^2 \langle u_d^2 \rangle} = e^{-\Delta K^2 c^0_d u_{sd}^2}.
\label{Eq: coherence residue}
\end{equation}

It is important to note that $u_{sd}^2$ varies depending on the material, serving as an indicator of the strength and reach of the displacement field around a defect. A higher $u_{sd}^2$ means atoms are more significantly displaced from their ideal positions, and the displacement field extends further from the defect. Consequently, a larger $u_{sd}^2$ results in greater depletion of coherent intensity when $c^0_d$ or momentum transfer $\Delta K$, and thus photon energy $h\nu$, are increased. The spread and magnitude of the displacement field typically depend on the elastic modulus of the material, with materials possessing a high elastic modulus tending to confine the distortion more tightly around the defect.
We therefore propose that Eq.~\ref{Eq: coherence residue} can be used to estimate the fraction of coherent intensity that originates from the static disorder within a crystalline sample.
To some extent, $F_{\text{C}}$ is equivalent to the DWF, where $c^0_d$ plays the role of an effective temperature.
Moreover, the contributions of static and thermal disorder to the total mean displacement can be approximated according to $\langle u_{tot}^2 \rangle = \langle u_T^2\rangle +\langle u_d^2 \rangle$. This is only possible if the thermal and static displacements are completely uncorrelated, for only then can we assume a vanishing cross-term $\langle \textbf{u}_T \cdot \textbf{u}_d \rangle$. As a result, the total portion of coherent intensity is given by the product of $F_{\text{C}}$ and DWF, namely $F_{\text{C}}(c_d)\text{DWF}(T) = e^{-\Delta K^2 \langle u_d^2 + u_T^2 \rangle}$.
Finally, we can make use of the DW theory to decompose the total ARPES spectral intensity $I$ as \cite{Hussain1980,White1986,Starnberg1993,Plucinski2008}:
\begin{multline}
    I(\textbf{k},E) = e^{-\Delta K^2 \langle u_d^2 + u_T^2 \rangle} I_{coh}(\textbf{k},E) + I_{inc}(\textbf{k},E,c_d,T)
    \label{total intensity}
\end{multline}
where $I_{coh}$ is the momentum-dependent coherent component that originates from Bloch electrons in the unperturbed crystal structure, $I_{inc}$ is the incoherent component associated electron-phonon, electron-defect scattering and intensity which results from photoelectrons scattered by other electrons.

\subsubsection{Finite mean free path effects}
\label{Results: Momentum broadening}
With mean free path effects, we refer to initial state effects, meaning effects that are intrinsic to the quantum states within the crystal and do not depend on the photoemission process. The mean free path $l$, not to be confused with the IMFP of the final states, represents the average distance traveled by an electron in the crystal before undergoing a scattering event. In this subsection, we focus on elastic scattering caused by static defects within a crystalline medium. If the static potential in the crystal deviates from the perfect periodic potential, then Bloch states undergo $\ket{n\textbf{k}} \rightarrow \ket{m\textbf{k}'}$ transitions, where $E_n(\textbf{k}) = E_m(\textbf{k}')$ is ensured by the absence of a time-dependence. Even though the scattering by defects is elastic, the finite probability of a scattering event implies a general reduction of the lifetime of $\ket{n\textbf{k}}$, which in turn, through the Heisenberg uncertainty principle, determines a broadening of the energy levels according to $\Delta E \sim \hbar/\tau$, and a mean free path of $l = v_{n\textbf{k}}\tau$.
Concretely, the one-electron spectral function measured by ARPES can be written in terms of the self-energy $\Sigma(\textbf{k},E)$ as
\begin{align}
I(\textbf{k},E) &\propto A(\textbf{k},E) \notag \\
&= -\frac{1}{\pi} \frac{\Im\Sigma(\textbf{k},E)}{[E-E_0(\textbf{k})-\Re\Sigma(\textbf{k},E)]^2+[\Im\Sigma(\textbf{k},E)]^2}.
\label{Eq: Self-energy}
\end{align}
The energy versus momentum dispersion relation is renormalized through the real part of the self-energy $\Re\Sigma(\textbf{k},E)$, while the imaginary part $\Im\Sigma(\textbf{k},E)$ gives some broadening to the spectral function. If, for practical purposes, we only consider static disorder, in the lowest-order Born approximation $\Re\Sigma(\textbf{k},E)\approx 0$ and $\Im\Sigma(\textbf{k},E)=1/2\tau_{n\textbf{k}}$.
Here, we argue that the mean free path of the electrons is geometrically limited by the average distance between defects, which in $d$ dimensions gives $l = c_d^{-1/d}$. As a consequence,
\begin{align}
\tau_{n\textbf{k}}^{-1} = c_d^{1/d} v_{n\textbf{k}}. \label{Eq: mfp}
\end{align}
Accordingly, the spectral function assumes a Lorentzian shape, with a FWHM given by $c_d^{1/d} v_{n\textbf{k}}$. Importantly, this implies that the present broadening mechanism affects all bands equally along momentum, regardless of their average velocity, but leads to different broadening along the energy axis. In particular, bands with higher velocities appear broader in energy than bands with smaller velocities.

\subsubsection{Inhomogeneous broadening}
\label{Results: Inhomogeneous broadening}

\begin{figure*}
\centering
\includegraphics[width=1\linewidth]{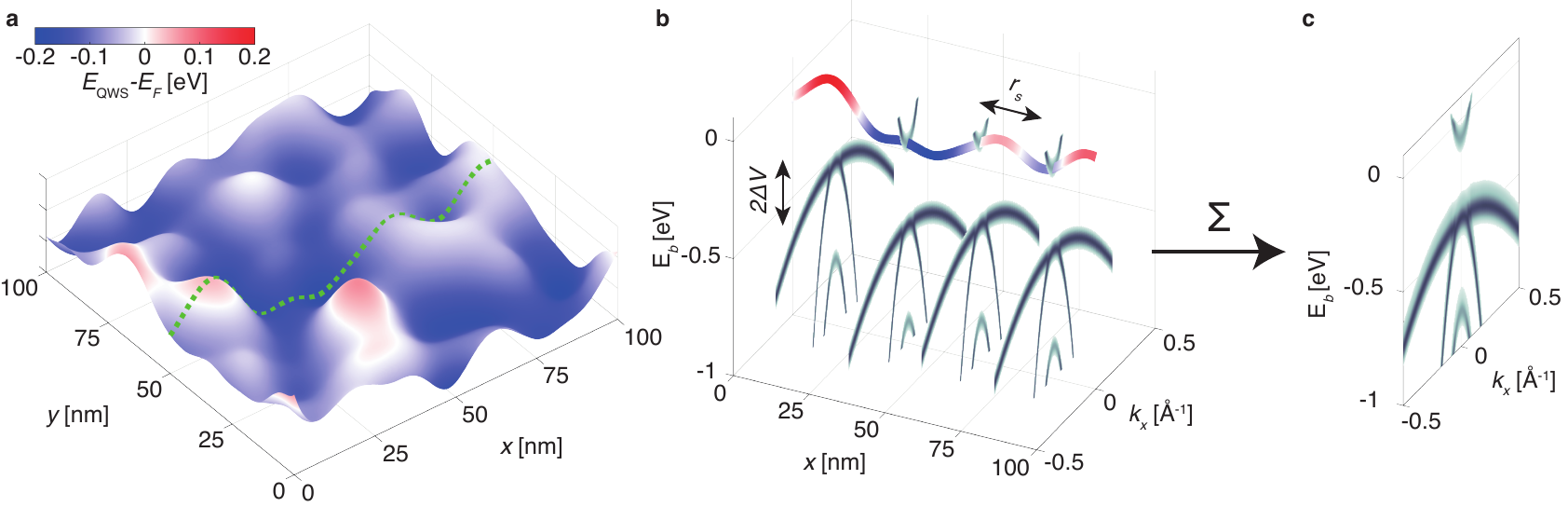}
\caption{(a) Simulated random potential, represented by the bottom of the QWS dispersion with $E_{\text{QWS}}=50$ meV, using a correlation length $r_s=6$ nm and a potential fluctuation strength of $\Delta V=150$ meV. (b) Schematic representation of an inhomogeneous potential, illustrated by the minimum of the QWS dispersion (green dotted line in (a)) along the $x$ direction. The band structure at selected points along $x$ is shown, emphasizing the varying offsets in the binding energy coordinate. Averaging over a large area of the crystal surface produces the inhomogeneously broadened spectrum shown in (c).}
\label{fig5}
\end{figure*}
Sputtering creates both interstitials and vacancies, which can act as dopants in InAs, and depending on the atomic species, become either positively or negatively charged \cite{Hglund2006}. The relatively small number of mobile carriers in the semiconductor can only partially screen the Coulomb potential generated by these defects. All of this results in variable band bending \cite{Schofield2013}, characterized by equipotential islands of typical size of the screening length $r_s$ and overall potential energy fluctuations of the order $\Delta V$, as schematically shown in Fig.~\ref{fig5}(a). The details of the shape of the islands depend on many factors, and are discussed in more detail in Ref.~\cite{Huang2022}. Here, we can estimate how $\Delta V$ depends on the defect concentration by considering the potential generated by a screened defect $\phi(r) = \pm \frac{e}{\kappa r}\exp\left(-\frac{r}{r_s}\right)$. Next, under the assumption that defects are statistically distributed within the crystal, the mean-squared value of the potential energy $\Delta V^2$ generated by $c_d$ defects is obtained from \cite{Skinner2014}:
\begin{equation}
\Delta V^2 = \int (e\phi(r))^2 c_d d^3r = \frac{2\pi e^4 c_d r_s}{\kappa} \propto r_s c_d.
\label{Eq: dV}
\end{equation}
 Where the screening length $r_s$ is much longer than the inter-defect distance, the potential at each point in space is equal to the sum of the potentials generated by the individual defects. By the central limit theorem, one can finally assume a Gaussian distribution of the potential values across the crystal with an energy scale $\Delta V \propto c_d^{1/2}$. Here, we neglect the potential dependence of $r_s$ on $c_d$, even though this should also vary as disorder is increased. Given that the beam spot size is several orders of magnitude larger than the typical screening lengths $r_s$, the photoemission spectra are the result of an average over all potential configurations, resulting in  inhomogeneous broadening, which convolves the pristine spectra with a Gaussian with FWHM of $2\sqrt{2\log{2}}\Delta V$ along the energy direction. This mechanism has been theoretically predicted by Chen et al. in Ref.~\cite{Chen2009}, where Coulomb disorder in the form of charged defects in an insulator leads to a Gaussian broadening of the ARPES spectra. They also predict, in a first approximation, a square root dependence of $\Delta V$ on the defect concentration.\\
Another relevant consequence of variable band bending, schematically shown in Fig.~\ref{fig5}, is that when the strength of the potential fluctuations $\Delta V$ reaches the same order of magnitude as the bandwidth of the QWS dispersion, it is inevitable that in certain islands of the crystal, the bottom of the QWS dispersion will lie above $E_F$, and therefore, the QWS pocket will not be locally occupied. As a consequence, the ARPES intensity of the QWS will be further depleted by this mechanism, as the regions where the QWS are absent do not contribute to the overall signal. 

To quantify this effect, we again appeal to the normal distribution of the fluctuating potential and calculate that, for a given fluctuation strength of the potential $\Delta V$ and the bandwidth of the QWS $E_{\text{QWS}}$, the probability that at random spot on the surface the QWS states are not populated by electrons is given by $\Phi(E_{\text{QWS}} / \Delta V)$, where $\Phi(x) = \frac{1}{\sqrt{2 \pi}}\int_{x}^{\infty}e^{-t^2/2}dt$ is the normal cumulative distribution function. Thus, the attenuation of the ARPES intensity of the QWS due to static disorder and variable band bending is given by:

\begin{equation}
I_{coh}^{\text{QWS}} \propto e^{-\Delta K ^2\langle u_d^2 \rangle} (1-\Phi(E_{\text{QWS}} / \Delta V)).
\label{Eq: QWS depletion}
\end{equation}
This formalism is equivalent to stating that whenever, in a certain region of the surface, the bottom of the QWS dispersion is above $E_F$, the contribution from this region to the ARPES signal of the QWS is zero. Conversely, when it is below $E_F$, the contribution is unity, independent of the local electron density.\\
Finally, we emphasize that the disordered potential discussed here is at the origin of the metal-insulator transition. It can be imagined that at a certain value $E_{\text{QWS}} / \Delta V \lesssim 1$, the electrons are confined in non-connected puddles, or in other words, they are below the percolation threshold, and their wave functions are exponentially localized. In this regime, conductivity at cryogenic temperatures follows the well-known variable-range hopping form of Efros-Shklovskii \cite{Shklovskii72, Shklovskii1984}. Also, as stated above, this effect is approximately scale-invariant, as it depends on the dimensionless parameter $E_{\text{QWS}} / \Delta V$. This suggests that the effect also plays a crucial role at much lower electron densities and fluctuation strengths, implying that observations made at high disorder levels can provide insight into what occurs below metal-insulator transitions for the 2D electron gases formed at surfaces.

\subsection{Depletion of coherent VB intensity}
\label{Results: Depletion of coherent VB intensity}

\begin{figure*}
\centering
\includegraphics[width=\linewidth]{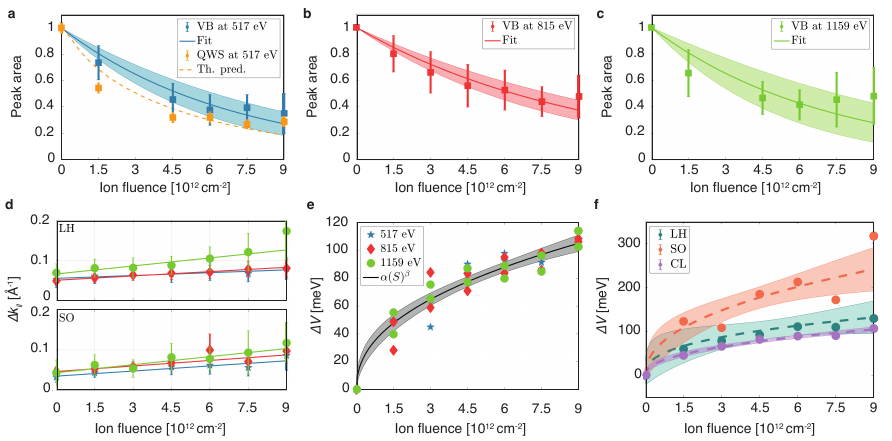} 
\caption{(a)–(c) Normalized peak areas of the valence bands as a function of Ar$^+$ dose at $ h\nu = 517$ eV, $815$ eV, and $1159$ eV, respectively. The continuous lines correspond to the optimized fitting function expressed in Eq.~\ref{Eq: FC fit}, and the colored areas represent the $95\%$ confidence intervals of the fitted parameters. Additionally, in (a), we show the intensity of the QWS extracted from the same spectrum measured at 517 eV, with the dashed line representing the theoretical prediction expressed in Eq.~\ref{Eq: QWS depletion}. (d) FWHM along $\Gamma$KX after subtracting the instrumental broadening versus Ar$^+$ dose for LH and SO bands, averaged over all MDCs for all three $h\nu$. (e) Change in the experimental FWHM of the In 4d core level, calculated through geometric subtraction of the FWHM at $D=0$ ions cm$^{-2}$. The black line represents an optimized power-law fitting function. (f) Comparison between $\Delta V$, obtained from the localized In 4d core states, and $\Delta k_{\parallel}$ extracted from the delocalized valence states after geometrically subtracting both $\Delta k_{\parallel}$ at $D=0$ ions cm$^{-2}$, multiplying by $ \frac{\partial E}{\partial k}$, and averaging across all $h\nu$.}
\label{fig6}
\end{figure*}

The experimental coherent intensity of the valence bands in the region $-4 \text{ eV}< E_b < -2 \text{ eV}$ as a function of disorder is obtained through the fitting procedure detailed in Sect.~\ref{Experimental procedures} and Appendix~\ref{App: coherence residue and linewidths}. The procedure is based on fitting MDCs within a $\pm 0.03$ eV integration window using pseudo-Voigt profiles, defined as a linear combination of a Gaussian and a Lorentzian function. For the valence bands, this approach provides direct access to the relationship between the coherence residue $F_{\text{C}}$ and the fluence of Ar$^+$ ions $D$, as depicted in Fig.~\ref{fig6} for various photon energies. To model the dependence of $F_{\text{C}}$ on $D$, we assume an average defect concentration that depends linearly on the Ar$^+$ dose $c^0_d = \eta D$, where $\eta$ represents the concentration rate constant. Utilizing the SRIM software, alongside the methodology described in Appendix~\ref{App: SRIM}, we derived estimates of $\eta$ for each of the three photon energies. The determined values of $\eta$ are listed in Tab.~\ref{tab1:fitting values}. Incorporating this into Eq.~\ref{Eq: coherence residue}, we derive the fitting function
\begin{equation}
    F_{\text{C}} = e^{-\Delta K^2 u_{sd}^2 c^0_d} = e^{-\Delta K^2 \gamma D},
    \label{Eq: FC fit}
\end{equation}

The parameter $\gamma \equiv \eta u_{sd}^2$ is optimized to align with the experimental observations. The optimal $\gamma$ values are listed in Tab.\ref{tab1:fitting values}, found in Appendix~\ref{App: coherence residue and linewidths}. We note that the $\gamma$ values are slightly higher for spectra measured with lower $h\nu$. This may indicate that the effective defect concentration, higher near the surface, is more strongly probed at lower $h\nu$, which, due to the IMFP, is more surface sensitive. This hypothesis is supported by the $\gamma$ value obtained from the exponential fit of the LEED Bragg peak area, which is significantly larger than all values measured with SX-ARPES.

Next, by multiplying $\gamma$ by $D$, we obtain the magnitude of the atomic displacement field squared. For instance, taking $\gamma = 10.7$ (measured at $h\nu = 517$ eV) and $D = 1.5 \times 10^{12} \text{ ions cm}^{-2}$ yields $\langle u_d^2 \rangle \approx 0.002 \text{ \AA}^2$. Taking the square root gives an average displacement of $0.04 \text{ \AA}$, which is below the independently obtained bound of  $\sim 0.1\text{ \AA}$ from the LEED data.  These displacement values are of the same order as the typical atomic displacement of $0.1 \text{ \AA}$ caused by thermal disorder at room temperature \cite{Shevchik1977}, further justifying the application of the Debye–Waller framework to this physical situation.

Finally, the simulated concentration rate constants $\eta$ enabled us to calculate the total displacement squared $u_{sd}^2$ resulting from the generation of a single defect. The computed $u_{sd}^2$ values are also documented in Tab.~\ref{tab1:fitting values}. Notably, the total squared displacement is more significant at $h\nu = 517 \text{ eV}$ compared to higher energies, which may again be attributed to the fact that photoelectrons excited with lower $h\nu$ predominantly probe regions of the crystal with the densest defect concentration, suggesting that the actual concentration rate constant $\eta$ might differ from the value determined using the SRIM software. Nonetheless, by computing a weighted average across all photon energies, the total squared displacement is quantified as $u_{sd}^2=0.46\pm 0.07 \text{ \AA}^2$.

\subsection{VB spectral broadening}
\label{Results: VB spectral broadening}
From the fitting procedure adopted to extract the area of the coherent peaks of the VB states, we could also determine the width of the MDC peaks for the SO and LH bands along the $\Gamma$KX direction as a function of the ion fluence $D$ for all $h\nu$. We deconvolved the instrumental broadening from the FWHM of the optimized pseudo-Voigt profiles by geometric subtraction. The results, displayed in Figures~\ref{fig6}(e), show $\Delta k_{\parallel}$, averaged over all MDCs, for the LH and SO bands as a function of $D$ for all $h\nu$. We employed weighted linear regression to determine the slope of $\Delta k_{\parallel}$ versus $D$. The optimized parameters are listed in Table~\ref{Tab: fwhm fitting values} in Appendix~\ref{App: coherence residue and linewidths}. For both bands, we observed an increase in linewidths along the momentum direction with increasing disorder for all $h\nu$. Notably, the $95\%$ confidence bounds are consistently positive, confirming the statistically significant trend of increasing linewidths with disorder for both LH and SO bands at all photon energies.

To corroborate the hypothesis that broadening due to the inhomogeneous potential plays a central role in this experiment, we fitted the In 4d core level measured under the same conditions using pseudo-Voigt functions. From the optimized functions, we extracted the FWHM. Since the determination of the Gaussian and Lorentzian contributions led to inconsistent results, we deconvolved the instrumental and lifetime broadening at $D=0$ cm$^{-2}$ using the equation $2\sqrt{2\log{2}}\Delta V = \sqrt{\text{FWHM}(D)^2-\text{FWHM}(0)^2}$. The resulting $\Delta V$ vs. $D$ is shown in Fig.~\ref{fig6}(e) for all $h\nu$. Based on the dependence of $\Delta V$ on $c_d$ shown in Eq.~\ref{Eq: dV}, we fitted all data points using the function $\alpha(D)^{\beta}$, where both $\alpha$ and $\beta$ have been optimized. The exponent $\beta_{\text{CL}} = 0.44 \pm 0.07$ is consistent with the expected square root dependence, providing further evidence of the broadening mechanism discussed in Sect.~\ref{Results: Inhomogeneous broadening}.
Finally, as mentioned in Sect.~\ref{Models: Effects of disorder on ARPES spectra}, to extract the magnitude of the variable band bending from the linewidths of the valence bands and compare it with that of core levels, we use $\Delta E = \frac{\partial E}{\partial k} \Delta k$, where  $\Delta k$ is obtained by geometrically subtracting the FWHM at $D=0$. The comparison is shown in Fig.~\ref{fig6}(f). From this figure, it is apparent that the broadening of the LH band, although slightly higher, quantitatively follows that of the In 4d core level. The SO band, on the other hand, has a significantly larger magnitude. In terms of the optimized parameters in the power-law fitting procedure, we obtained for the LH band $\beta_{\text{LH}} = 0.38 \pm 0.33$, while for the SO band, we get $\beta_{\text{SO}} = 0.45 \pm 0.19$. We expect both the LH and SO bands to exhibit broader linewidths due to the finite mean free path mechanism discussed in Sect.~\ref{Results: Momentum broadening}, which is absent for the localized core states. In particular, we anticipate a larger discrepancy for the SO band due to its higher velocity, $v_{\text{SO}} = 8.53$ eV\AA$\hbar^{-1}$, in the region of interest, compared to $v_{\text{LH}} = 4.49$ eV\AA$\hbar^{-1}$. If, from the VB data shown in Fig.\ref{fig6}(f), we geometrically subtract the $\Delta V$ obtained from the core levels and divide the result by the respective velocities, the LH and SO data once again show good agreement, with a remnant broadening of the SO band with respect to LH. Furthermore, as shown in Fig.~\ref{figA1} of Appendix~\ref{App: mfp}, converting these results into an elastic mean free path yields very good agreement with the inter-defect distance observed via STM in Fig.\ref{fig3}(c).

After subtracting the effects of the inhomogeneous potential and finite mean free path from the VB lineshapes, the remaining broadening of the SO states suggests that additional mechanisms may be involved, mechanisms that are not as pronounced for the LH band. A plausible explanation is that static disorder induces spatial inhomogeneities in the spin-orbit coupling across the sample. At the $\Gamma$-point, and in the absence of disorder, the spin-orbit splitting is $\Delta_{\text{SO}} = 0.41$ eV. Without spin-orbit coupling, the HH, LH, and SO bands would be degenerate at $\Gamma$. This highlights how variations in spin-orbit coupling could produce substantial fluctuations in the energy position of the SO band, potentially explaining the remnant difference in linewidths.

We also make a final remark about the method used to extract inhomogeneous broadening. While extracting potential fluctuations from MDCs and then converting them into energy broadening may not seem the most direct approach, e.g., directly extracting the linewidths along energy at points in \textbf{k}-space where the bands have zero velocity appears more straightforward, we argue that our method ensures greater robustness against artifacts. Given the increase in a DOS-like background relative to the coherent component of the spectra, measuring the linewidths along momentum guarantees that any potential distortion of the lineshapes due to this effect is ruled out, since the DOS-like background remains a mere constant along \textbf{k}. On the other hand, when measuring the linewidth of extrema under these conditions, the proper disentanglement of coherent and incoherent intensity becomes technically more challenging, and the risk of dealing with artifacts rather than physical phenomena increases. However, for completeness, we still show EDCs measured at the $\Gamma$-point in Fig.\ref{figA4}(a) of Appendix~\ref{App: EDCs at G}, which corroborate both the inhomogeneous potential picture and the change in spin-orbit coupling upon increasing disorder.

\subsection{Depletion of coherent QWS intensity}
\label{Results: Depletion of QWS intensity}
As discussed above, the depletion of ARPES intensity originating from QWS due to disorder is a more complicated topic than the depletion of the VB ARPES intensity. While for the latter, the only mechanism responsible for the attenuation of the magnitude of the quasiparticle peaks is a Debye–Waller-like mechanism, for the QWS ARPES intensity, this effect is present along with additional inhomogeneous potential effects. 
The ARPES intensity from the QWS has been obtained as explained in Sect.~\ref{Experimental procedures} and is shown in Fig.~\ref{fig6}(a), together with the intensity originating from the valence bands extracted from the same spectrum. This figure already demonstrates that the intensity of the QWS is significantly more suppressed than that of the VB. 

Eq.~\ref{Eq: QWS depletion}, is an expression that, apart from the coherence residue, depends on a single parameter, namely the ratio $E_{\text{QWS}}/\Delta V$. ARPES directly accesses  $E_{\text{QWS}}$, which in our case is approximately 50 meV. For $\Delta V$, we use the fitting function optimized on the core level data. The combination with the coherence residue extracted from the VB yields the dashed line in Fig.~\ref{fig6}(a), which accurately describes the dependence of the QWS intensity on disorder. Finally, we point out that this analysis assumes that $E_{\text{QWS}}$ is not affected by disorder. In principle, a corollary of the effect demonstrated here, and schematically shown in Fig.~\ref{fig5}(a), is that the space available to electrons is reduced upon increase of the disorder. In particular, at constant electron density, this implies a smaller average distance between electrons, and therefore larger Coulomb repulsion. This could result in a decrease in $E_{\text{QWS}}$. As a result, the ratio $E_{\text{QWS}}/\Delta V$ changes, and the 2D electron system would localize faster.

\section{Conclusions}
We have developed a framework to explain two of the most common observations in semiconductor photoemission studies: low coherent intensity and broadened lineshapes. For the former, we extended the Debye–Waller factor, originally developed to describe thermal disorder in diffraction and photoemission experiments, to account for static disorder in ARPES measurements. Our experimental results, obtained on ion-bombarded InAs(110), exhibit quantitative agreement with the proposed theory and diffraction data, thereby reinforcing its validity in explaining the depletion of coherent intensity in the valence and conduction bands.

Our measurements show that suppression of the QWS intensity exceeds that of the valence bands, and that the linewidths of valence band quasiparticle peaks increase with disorder concentration. These observations support our description that charged defects generate regions of varying potential within the semiconductor. In the first case, disorder-induced variable band bending creates regions where QWS are unoccupied, leading to an overall reduction in signal from these states relative to the valence bands. In the second case, averaging over large regions of the crystal, namely over lengths exceeding the screening length, results in Gaussian broadening of the lineshapes along the energy axis. This effect is further enhanced by the limited mean free path of the initial states, which contributes additional Lorentzian broadening. For both localized core levels and delocalized valence states, we consistently extracted the magnitude of potential energy fluctuations and observed a disorder dependence consistent with theoretical predictions.

Overall, our observations on ion-bombarded InAs(110) provide valuable insights into the universal characteristics of ARPES intensity depletion and spectral broadening in semiconductors and insulators. Specifically, they clarify the role of the static Debye–Waller mechanism in suppressing coherent peaks and highlight the impact of potential fluctuations caused by charged defects, an effect that plays a crucial role in driving key phenomena such as the metal–insulator transition.

\section*{Acknowledgements}
The authors thank M. Heinrich for his support during the STM experiment at the PEARL beamline of the Swiss Light Source, Paul Scherrer Institut. Additionally, they acknowledge L. Nue and P. Aescher for their technical support during the experiments. P.C. was supported by the Microsoft Corporation.

\appendix
\renewcommand{\thefigure}{A\arabic{figure}}
\setcounter{figure}{0}

\section{LEED}
\label{App: leed}
LEED patterns were measured on freshly cleaved InAs(110) surfaces before and after Ar$^+$ sputtering at an electron kinetic energy of 100 eV. Cleavage was performed at liquid nitrogen temperature, while sputtering was carried out using parameters equal to those described in Sect.~\ref{Experimental procedures}, namely an Ar$^+$ ion kinetic energy of 3 keV and an Ar partial pressure of $5 \times 10^{-8}$ mbar. Sputtering was conducted in cycles of 20 s each ($D = 1.5 \times 10^{12}$ cm$^{-2}$), until a total ion dose of $D = 7.5 \times 10^{12}$ cm$^{-2}$ was reached. The results for the three different defect concentrations are presented in Fig.~\ref{fig4}.
 
\section{Determination of mean free path}
\label{App: mfp}
\begin{figure}
\centering
\includegraphics[width=0.9\linewidth]{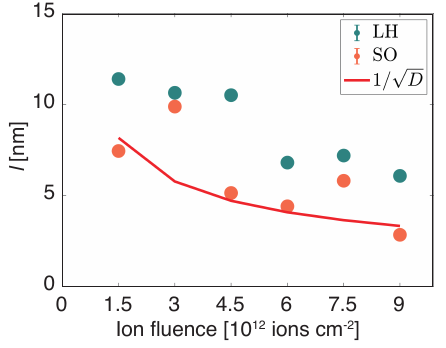}
\caption{Mean free path extracted from the valence bands after disentangling the contribution from the fluctuating potential. Green circles represent results from the LH band, orange circles correspond to the SO band, and the red line indicates the inter-defect distance estimated from STM data.}
\label{figA1}
\end{figure}

The fact that both mean free path effects and inhomogeneous broadening contribute to the VB lineshapes, combined with the availability of inhomogeneous broadening estimates from the core levels, allows us to disentangle these two contributions and reconstruct the mean free path of the VB states. Specifically, we geometrically subtract the $\Delta V$ measured from the core levels from $\frac{\partial E}{\partial k}\Delta k$, where $\Delta k$ is obtained by geometrically subtracting the FWHM at $D = 0$. According to Eq.~\ref{Eq: mfp}, the result corresponds to the inverse lifetime of the VB states. To obtain the mean free path, we divide this quantity by the band velocity and take the reciprocal.
In Fig.\ref{figA1}, we present the resulting mean free path extracted from both the LH and SO bands as a function of ion dose. When compared with the inter-defect distance measured by STM in Fig.\ref{fig2}, the extracted values show excellent agreement for the SO band, while the LH band exhibits a discrepancy of approximately 30\%. This discrepancy may arise from the fact that quasiparticle peaks are best described by Voigt functions, and using the geometric approximation to disentangle the Lorentzian and Gaussian components of the FWHM can introduce inaccuracies. An additional contributing factor could be slight variations in the sputtering process: although the same parameters were used in both the STM and SX-ARPES experiments, the sputtering guns were different, potentially leading to a mismatch between the nominal ion doses and the effective doses experienced during SX-ARPES.

\section{Depth distribution of defects}
\label{App: SRIM}
\begin{figure}
\centering
\includegraphics[width=0.85\linewidth]{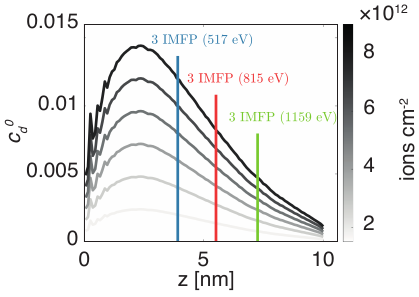}
\caption{Concentration of In and As vacancies as a function of depth, shown at different Ar$^+$ beam fluences. The colored vertical lines show the probing depths of the three photon energies in InAs. These results were obtained by running a simulation with the SRIM software package \cite{Ziegler2010}.}
\label{figA2}
\end{figure}

Fig.~\ref{figA2} shows the out-of-plane distribution of vacant In and As sites calculated using SRIM software \cite{Ziegler2010}. Each of these displaced atoms stops at some interstitial site, effectively creating a second defect. 
For comparison purposes, the probing depth in the pristine sample, namely three times the IMFP \cite{Strocov2019}, is plotted on top of the defect distribution. Since, according to \cite{Aukerman1968}, the energy to displace an In atom is slightly smaller compared to that of As, the In atoms are displaced more frequently during the bombardment process. Furthermore, the peak position at $\sim 20 \text{ \AA}$ is explained by the cascade effect, generated by the Ar ions that displace target atoms, which in turn create more defects as they propagate away from the surface. Ref.~\cite{Chen2021} has calculated that 63 \% of the sputter induce defects are cancelled within a few ps. Here, to take care of this significant effect we multiply the total number of displaced atom by 0.37. We point out that the simulation does not include temperature effects, which lead to self-annealing processes and therefore to a general reduction of the defects' concentration.
Further, the simulation does not take into account the fact that atoms sitting at damaged areas of the crystal need less energy to be displaced. 
This leads to an underestimate of the actual concentration of defects. Finally, the quantity $c_d(z)$ shown in Fig.~\ref{figA2} allows us to estimate the effective concentration of vacancies perceived by photoelectrons with different probing depths. This is achieved through weighted averaging of $c_d$ over the depth coordinate, with the weight being $\exp(-z/\text{IMFP}(h\nu))$.

\section{Role of vacancies and interstitials in ARPES}
\label{App: coherence residue}
\begin{figure}
\centering
\includegraphics[width=0.8\linewidth]{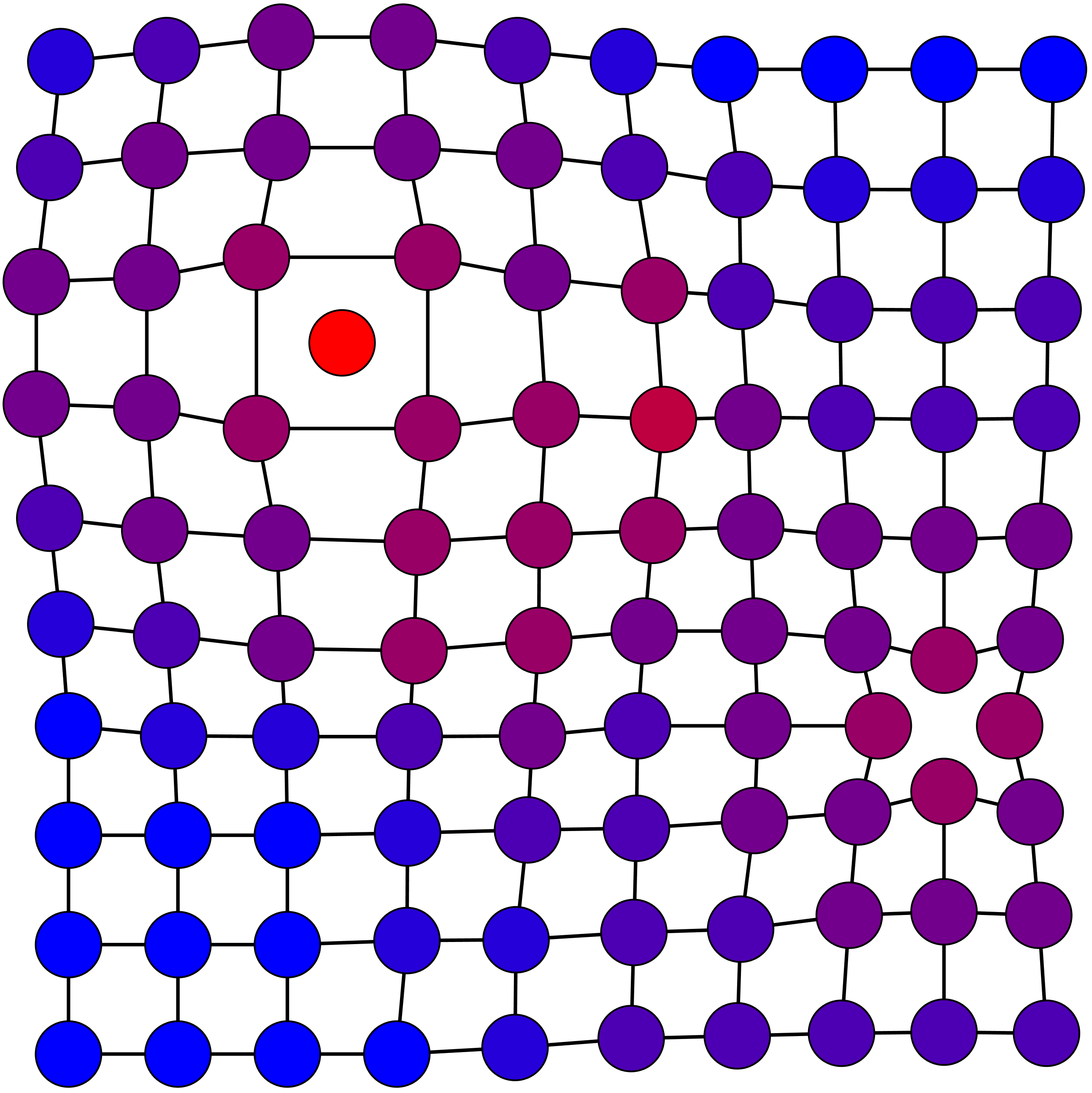}
\caption{Schematic representation of a crystal lattice following the creation of a vacancy and an interstitial defect. Atoms deviating from their equilibrium lattice positions are color-coded to reflect the extent of their displacement, with a color gradient indicating the magnitude. Note that actual atomic displacements caused by point defects occur on a length scale typically two orders of magnitude smaller than the interatomic distance. This visualization aids in quantifying the mean square displacement, $\langle u_d^2 \rangle$, highlighting secondary displacement effects induced by defect formation.}
\label{figA3}
\end{figure}
Here, we demonstrate that the mere presence of vacancies and interstitials, in absence of secondary displacements, cannot account for the significant depletion of coherent intensity observed experimentally. To elucidate this, we commence from the foundational principles, adopting an initial state $\Psi_{\textbf{k}_i} = \sum_m e^{i\textbf{k}_i\cdot \textbf{R}_m}\phi_{\textbf{k}_i}(\textbf{r}-\textbf{R}_m)$. We proceed by considering plane waves as final states $\Psi_{\textbf{k}_f}(\textbf{r}) = e^{i\textbf{k}_f \cdot \textbf{r}}$. It has been established that plane wave final states are not an indispensable prerequisite for achieving our ultimate result. This assertion is supported by the work of Shevchik, which utilized augmented plane waves instead. Consequently, the ARPES coherent intensity $I_{coh}$ is expressed as 
\begin{align*}
I_{coh} &= |\bra{\Psi_{\textbf{k}_f}} \textbf{A} \cdot \textbf{p} \ket{\Psi_{\textbf{k}_i}}|^2 \propto |\int d^3r \: e^{-i \textbf{k}_f\cdot \textbf{r}} \Psi_{\textbf{k}_i}(\textbf{r})|^2 =\\
&= | \sum_m e^{i(\textbf{k}_i-\textbf{k}_f)\cdot \textbf{R}_m}\int d^3r \: \phi_{\textbf{k}_i}(\textbf{r}) e^{i\textbf{k}_f \cdot \textbf{r}}|^2 = \\
&=S(\textbf{k}_i-\textbf{k}_f) |\mathcal{F}(\phi_{\textbf{k}_i}(\textbf{k}))|^2,
\end{align*}

where $S(\textbf{k}) = |\sum_m e^{i\textbf{k}\cdot \textbf{R}_m}|^2$. We proceed by making several assumptions regarding the static disorder within our crystal. It is presumed that atoms initially positioned at lattice sites $\textbf{R}_{i_1}, \textbf{R}_{i_2},...,\textbf{R}_{i_M}$ are permanently displaced from their ideal locations. The displacement vectors are denoted as $\textbf{U}_{i_1}, \textbf{U}_{i_2},...,\textbf{U}_{i_M}$. It is important to note that all other atoms $m \notin I={i_1,i_2,...,i_M}$ remain at their ideal positions. We then endeavor to compute $S(\Delta \textbf{k})$ for the disordered crystal, where $\Delta \textbf{k} = \textbf{k}_i-\textbf{k}_f$ is used for brevity. For a specific configuration of disordered atoms $I$, corresponding to a disorder concentration $\frac{M}{N}$, where $N$ represents the total number of atoms, $S(\Delta \textbf{k}) = | \sum_{m \notin I}e^{i \Delta \textbf{k}\cdot \textbf{R}_m} + \sum_{n\in I} e^{i \Delta \textbf{k}(\textbf{R}_n+\textbf{U}_n)}|^2$. In systems of considerable size, implying $N \gg 1$, Kohn and Luttinger, Ref.~\cite{Kohn1957}, have illustrated that quantities such as $S(\Delta \textbf{k})$ closely approximate their ensemble average for the majority of microscopic configurations. Thus, we can approximate $S(\Delta \textbf{k})\approx \langle S(\Delta \textbf{k}) \rangle$ for sufficiently large values of $N$. Subsequently, we calculate the ensemble average as follows:
\begin{widetext}
\begin{equation}
    \langle S(\Delta \textbf{k}) \rangle = \sum_{I} P_I \int d^3 U_{i_1}
    f(\textbf{U}_{i_1})\int d^3 U_{i_2}f(\textbf{U}_{i_2})...\int d^3 U_{i_M} f(\textbf{U}_{i_M})\left| \sum_{m \notin I}e^{i \Delta \textbf{k}\cdot \textbf{R}_m} + \sum_{n\in I} e^{i \Delta \textbf{k}(\textbf{R}_n+\textbf{U}_n)}\right|^2,
\end{equation}
\end{widetext}
where the initial summation encompasses all conceivable $N \choose M$ configurations, and the integrals $\int d^3 U_{i}f(\textbf{U}_{i})$ span the entire volume of the crystal. The function $f(\textbf{U}_i)d^3U_i$ signifies the probability of locating the displaced atom within the infinitesimal volume $d^3U_i$ centered around the position $\textbf{R}_i+\textbf{U}_i$. We adopt the premise that the probability density function $f$ is uniform for every atom and remains unaffected by the displacement of other atoms. The probability $P_I$ for a single configuration equates to the reciprocal of the total number of configurations, expressed as $P_I = \left[ N \choose M \right]^{-1}$.
By expanding the ensemble average, we derive the following expression:
\begin{widetext}
\begin{equation}
    \langle S(\Delta \textbf{k}) \rangle = \underbrace{\sum_I P_I\sum_{\substack{m \notin I\\n \notin I}} e^{i \Delta \textbf{k}\cdot (\textbf{R}_m-\textbf{R}_n)}}_{\text{I}} +  \underbrace{2\Re\left[ f(\Delta \textbf{k}) \sum_I P_I\sum_{\substack{m\notin I\\n \in I}} e^{i \Delta\textbf{k}\cdot(\textbf{R}_m -\textbf{R}_n)} \right]}_{\text{II}}
    + \underbrace{|f(\Delta \textbf{k})|^2 \sum_I P_I\sum_{\substack{m \in I\\n \in I}} e^{i \Delta \textbf{k}\cdot (\textbf{R}_m-\textbf{R}_n)}}_{\text{III}},
\end{equation}
\end{widetext}
where $f(\Delta \textbf{k}) = \int d^3U f(\textbf{U})e^{i \Delta \textbf{k} \cdot \textbf{U}}$ represents the Fourier transform of the probability density function $f$. Ultimately, it is feasible to demonstrate that the term I can be reformulated into $\frac{(N-M)^2}{N^2}\sum_{m,n}e^{i \Delta \textbf{k}\cdot (\textbf{R}_m-\textbf{R}_n)}=\frac{(N-M)^2}{N^2}\left| \sum_m e^{i \Delta \textbf{k}\cdot \textbf{R}_m} \right|^2$. Similarly, it can be shown that term II equates to $2\Re(f)\frac{(N-M)M}{N^2}\left|^2 \sum_m e^{i \Delta \textbf{k}\cdot \textbf{R}_m} \right|$, and term III corresponds to $|f|^2\frac{M^2}{N^2}\left| \sum_m e^{i \Delta \textbf{k}\cdot \textbf{R}_m} \right|^2$. By amalgamating these terms, the entire expression can be simplified to:
\begin{align}
    \langle S(\Delta \textbf{k}) \rangle &= \left| 1-(1-f(\Delta \textbf{k}))\frac{M}{N} \right|^2 \left|\sum_{m} e^{i \Delta \textbf{k}\cdot \textbf{R}_m}\right|^2.
\end{align}
For homogeneous probability distributions $f$, $\langle S(\Delta \textbf{k}) \rangle$ is proportional to the interference function of the perfect crystal $S(\Delta \textbf{k})$, with the factor being $\left| 1-\frac{M}{N} \right|^2\leq 1$.

\section{Inhomogeneous broadening at the $\Gamma$-point}
\label{App: EDCs at G}
\begin{figure}
\centering
\includegraphics[width=0.8\linewidth]{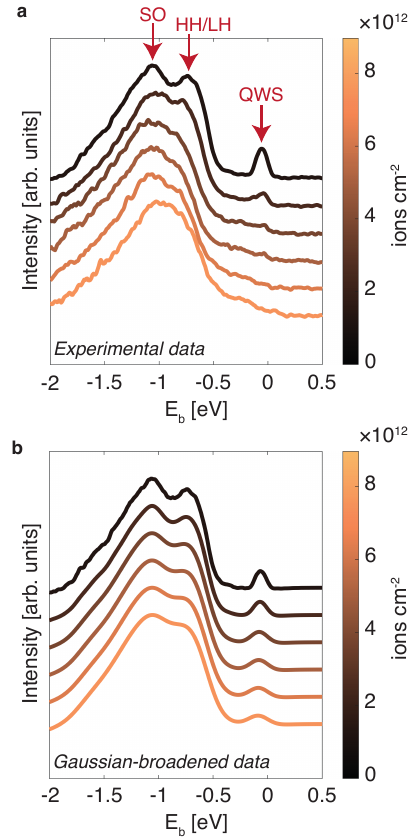}
\caption{(a) Normalized EDC extracted at the $\Gamma$-point with an integration window of $\pm 0.03$ \AA$^{-1}$ as a function of Ar$^+$ ion fluence, after subtracting an EDC extracted from the same spectrum at the K-point, and plotted with a vertical offset at the zero position of the intensity. In the figure, we also identify the SO band, the HH and LH bands, which are degenerate at $\Gamma$, and the QWS. (b) The EDC shown in black in (a) is Gaussian broadened using $\Delta V$ measured from the In 4d core levels.}
\label{figA4}
\end{figure}

To further corroborate the inhomogeneous picture described in this paper, we extracted EDCs at the $\Gamma$-point with an integration window of $\pm 0.03$ \AA$^{-1}$, after subtracting an EDC extracted from the same spectrum at the K-point, and plotted them as a function of Ar$^+$ fluence in Fig.~\ref{figA4}(a). The subtraction of the EDC at the K-point allows us to remove the DOS-like incoherent background. In this figure, we also identify the three main peaks, which correspond to the SO band, separated from the degenerate HH and LH bands by $0.41$ eV, and the QWS. The figure shows that with increasing disorder, the splitting between the SO and HH/LH bands becomes less and less visible. This corroborates the inhomogeneous broadening picture. We also observe the merging of the two peaks, which suggests that the spin-orbit interaction is likely weakened by the disorder. In Fig.\ref{figA4}(b), we take the EDC at $D=0$ cm$^{-1}$ and Gaussian broaden it using $\Delta V$ obtained from the In 4d core level fitting. This shows the expected decrease in resolution between the two peaks; however, it also suggests that the spin-orbit coupling likely needs to change to fully reproduce the evolution shown in Fig.\ref{figA4}(a).

\section{Separation of coherent and incoherent intensity}
\label{App: background subtraction}
To separate the coherent peaks from the incoherent background we smoothed the spectral data along the momentum ($\textbf{k}$) axis using Gaussian convolution. Following this, we calculated the second derivative of the spectrum along $\textbf{k}$, identifying the peak bases at points where the second derivative exceeded a specific threshold. To capture the full extent of the coherent peaks, including their tails, we expanded the peak base by a fixed constant. We then employed linear interpolation across the peak bases to isolate the coherent intensity from the incoherent background within the peak region.

\begin{figure*}
\centering
\includegraphics[width=\linewidth]{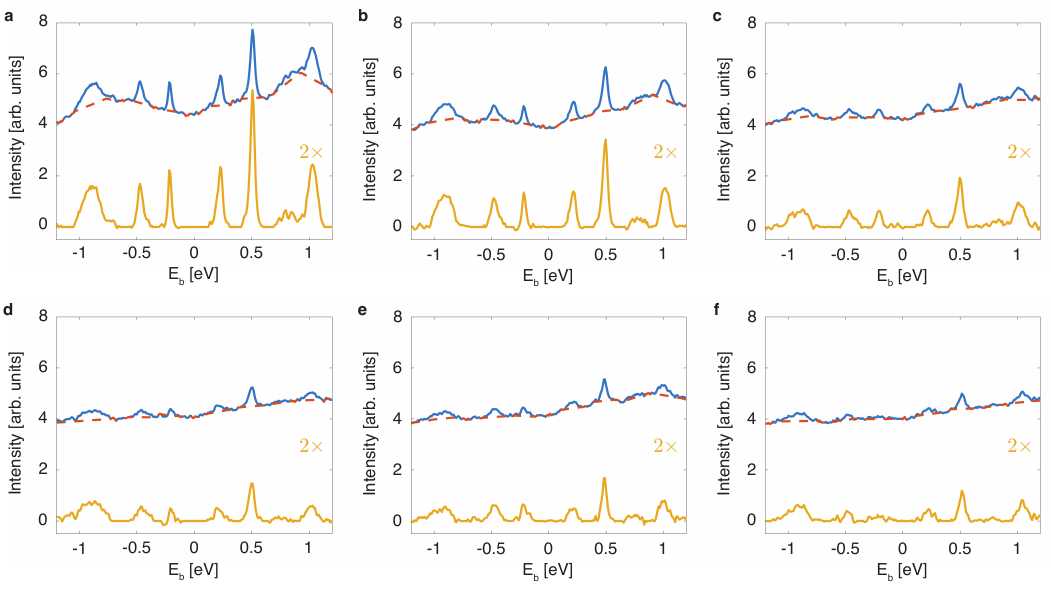} 
\caption{Separation of the coherent and incoherent signal from the dataset measured at $h\nu = 517 \text{ eV}$. (a)-(f) show the MDC extracted at $E_b = -2.63 \text{ eV}$ for all ion doses in increasing order. For each figure, the blue solid line represents the total intensity, the red dashed line is the background and the yellow solid line is the coherent intensity.}
\label{figA5}
\end{figure*}

\section{Extraction of the coherence residue and linewidths from ARPES data}
\label{App: coherence residue and linewidths}
In the appendix, we explain the process of extracting the coherence residue and linewidths from ARPES data. Initially, we extract the coherent intensity from the ARPES data using the background subtraction method outlined in Appendix~\ref{App: background subtraction}. For each dataset measured at three different photon energies, we analyze $10$ MDCs with a $\pm 0.03 \text{ eV}$ integration window in the valence band region $-3.5 \text{ eV} < E_b < -2.5 \text{ eV}$. We fit the four peaks, corresponding to the LH and SO bands, with pseudo-Voigt profiles, defined as a linear combination of Gaussian and Lorentzian components. The fit parameters include the peak height, FWHM, mixing ratio between Gaussian and Lorentzian components, and the $k_x$ position. Each peak's area is normalized to its area at zero Ar$^+$ dose for comparison.

Subsequently, we calculate the average normalized peak areas of the LH and SO bands across all binding energies for each Ar$^+$ beam fluence. This average provides the data point for each dose, with the standard deviation serving as the uncertainty estimate for each point. The final step involves presenting the averaged intensities and FWHM values for the LH and SO bands as a function of Ar$^+$ dose, as depicted in Fig.~\ref{fig6}. Our method allows us to systematically assess the impact of sputtering-induced disorder on the ARPES-measured electronic structure.

The optimal values $\gamma$, obtained by fitting $F_{\text{C}}$ vs. $D$ with Eq.~\ref{Eq: FC fit}, are displayed in the third column of Tab.~\ref{tab1:fitting values}. The fourth column shows the concentration rate constant $\eta$ calculated with the SRIM software, while the fifth column contains $u_{sd}^2$.
\begin{table}[b]
\caption{\label{tab1:fitting values}%
Optimal values for the parameter $\gamma$ expressed in Eq.~\ref{Eq: FC fit} for the three photon energies used in the experiment.
}
\begin{ruledtabular}
\begin{tabular}{ccccc}
\textrm{Method}&
\textrm{$h \nu /E_{kin}$ [eV]}&
\textrm{$\gamma$ [\AA$^{4}]$}&
\textrm{$\eta$  [cm$^{2}$]}&
\textrm{$u_{sd}^2$ [\AA$^{2}$]}\\
\colrule
SX-ARPES & $517$  & $10.7 \pm 2.7$ &$1.20\times 10^{-15}$&  $0.89 \pm 0.23$\\
SX-ARPES &$815$ & $5.1 \pm 1.0$ & $1.22\times 10^{-15}$ &  $0.42 \pm 0.08$\\
SX-ARPES &$1159$  & $4.6 \pm 1.9$ & $1.21\times 10^{-15}$&  $0.38\pm 0.16$\\
LEED &$100$ & $81.6 \pm 9.5$ & –  &  –\\
\end{tabular}
\end{ruledtabular}
\end{table}

\begin{table}[b]
\caption{\label{Tab: fwhm fitting values}%
Optimal values for the parameters $a$ and $b$ obtained by fitting $\Delta k_{\parallel}$ as a function of $D$ using $\Delta k_{\parallel} = a D +b$, as shown in Figs.~\ref{fig6}(e) and (f).
}
\begin{ruledtabular}
\begin{tabular}{cccc}
\textrm{Band}&
\textrm{$h \nu $ [eV]}&
\textrm{$a$  [\AA]}&
\textrm{$b$ [$10^{-3}$\AA$^{-1}$]}\\
\colrule
\multirow{3}{*}{\rotatebox[origin=c]{0}{LH}} & $517$ & $23 \pm 15$ & $56\pm 8$\\
&$815$  & $36 \pm 7$ & $51 \pm 3$\\
&$1159$ & $66 \pm 40$ & $67 \pm 16$\\
\colrule
\multirow{3}{*}{\rotatebox[origin=c]{0}{SO}} & $517$ & $42 \pm 24$ & $35\pm 10$\\
&$815$  & $46\pm 23$ & $46 \pm 9$\\
&$1159$ & $67\pm 39$ & $43 \pm 16$\\
\end{tabular}
\end{ruledtabular}
\end{table}

\begin{table}[b]
\caption{\label{tab3:fitting values}%
Optimal values for the power-law fitting of $\Delta V \propto (D)^{\beta}$.
}
\begin{ruledtabular}
\begin{tabular}{ccc}
\textrm{Band}&
\textrm{$\beta$}\\
\colrule
In 4d & $0.44 \pm 0.07$\\
LH & $0.38 \pm 0.33$\\
SO & $0.45 \pm 0.19$\\
\end{tabular}
\end{ruledtabular}
\end{table}

\section{XPS fitting}
\label{App: XPS fitting}
\begin{figure*}
\centering
\includegraphics[width=\linewidth]{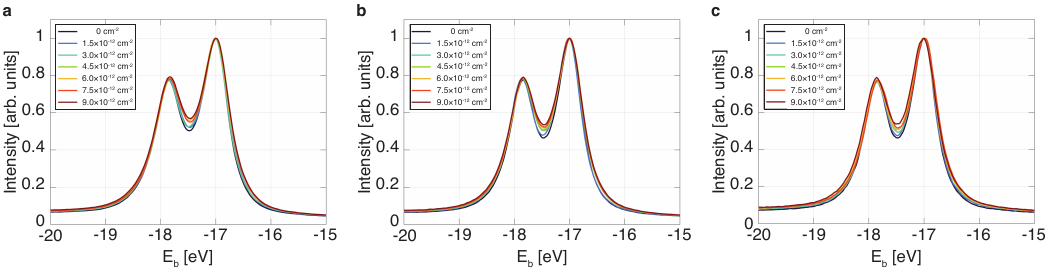} 
\caption{In 4d core levels measured at (a) $517$ eV, (b) $815$ eV and (c) $1159$ eV for all Ar$^+$ fluences.}
\label{figA6}
\end{figure*}

To extract the FWHM for the In 4d core states, we first normalized the data to the height of the In 4d$_{5/2}$ peak. The peaks are shown for all three $h\nu$ in Fig.~\ref{figA6}. All spectra are fitted using pseudo-Voigt functions. Since a priori knowledge of how disorder affects the spin-orbit states is unavailable, we opted for a parameter-independent fitting procedure for the two peaks shown in Fig.~\ref{figA6}. More precisely, we did not constrain the relative height, the distance, the mixing ratio, or the FWHM of the two peaks. These were all treated as independent parameters to be optimized. At the same time, we account for the effect of inelastic electron scattering by performing a Shirley background subtraction \cite{Shirley1972}, which was optimized self-consistently with the peak fitting. By observing the intensity between the peaks, it is already possible to deduce an increasing trend in the FWHM.
\bibliography{MyRefs}

\end{document}